\documentclass{article}

\usepackage{PRIMEarxiv}

\usepackage[utf8]{inputenc} 
\usepackage[T1]{fontenc}    
\usepackage{hyperref}       
\usepackage{url}            
\usepackage{booktabs}       
\usepackage{amsfonts}       
\usepackage{nicefrac}       
\usepackage{microtype}      
\usepackage{lipsum}
\usepackage{fancyhdr}       
\usepackage{graphicx}       
\graphicspath{{media/}}     

\usepackage{bm}
\usepackage{adjustbox}

\title{Lightweight High-Performance Blind Image \\
Quality Assessment}

\author{
  Zhanxuan Mei\textsuperscript{\textsection} \\
  University of Southern California \\
  Los Angeles, USA\\
  \texttt{zhanxuan@usc.edu} \\
   \And
  Yun-Cheng Wang\textsuperscript{\textsection} \\
  University of Southern California \\
  Los Angeles, USA\\
  \texttt{yunchenw@@usc.edu} \\
   \And
  Xingze He \\
  Meta Platform, Inc. \\
  Menlo Park, California, USA\\
  \texttt{xingze.he@fb.com} \\
   \And
   Yong Yan \\
  Meta Platform, Inc. \\
  Menlo Park, California, USA\\
  \texttt{yyan22@fb.com} \\
   \And
  C.-C. Jay Kuo \\
  University of Southern California \\
  Los Angeles, USA\\
  \texttt{cckuo@sipi.usc.edu} \\
}

\begin{document}
\maketitle
\begingroup\renewcommand\thefootnote{\textsection}
\footnotetext{Equal contribution}
\endgroup

\begin{abstract}
Blind image quality assessment (BIQA) is a task that predicts the
perceptual quality of an image without its reference.  Research on BIQA
attracts growing attention due to the increasing amount of
user-generated images and emerging mobile applications where reference
images are unavailable.  The problem is challenging due to the wide
range of content and mixed distortion types.  Many existing BIQA methods
use deep neural networks (DNNs) to achieve high performance.  However,
their large model sizes hinder their applicability to edge or mobile
devices.  To meet the need, a novel BIQA method with a small model, low
computational complexity, and high performance is proposed and named
``GreenBIQA" in this work. GreenBIQA includes five steps: 1) image
cropping, 2) unsupervised representation generation, 3) supervised
feature selection, 4) distortion-specific prediction, and 5) regression
and decision ensemble. Experimental results show that the performance of
GreenBIQA is comparable with that of state-of-the-art deep-learning (DL)
solutions while demanding a much smaller model size and significantly
lower computational complexity. 
\end{abstract}


\section{Introduction}\label{G_sec:introduction}

{O}{bjective} image quality assessment (IQA) can be
classified into three categories: full-reference IQA (FR-IQA),
reduced-reference IQA (RR-IQA), and no-reference IQA (NR-IQA). FR-IQA
methods evaluate the quality of images by comparing distorted images
with their reference images.  Quite a few image quality metrics such as
PSNR, SSIM~\cite{wang2004image}, FSIM~\cite{zhang2011fsim}, and MMF
\cite{liu2012image} have been proposed in the last two decades.  RR-IQA
methods (e.g.  RR-SSIM~\cite{rehman2012reduced}) utilize part of
information from reference images to evaluate the quality of underlying
images. RR-IQA is more flexible than FR-IQA. NR-IQA, also called blind
image quality assessment (BIQA), is needed in two scenarios. First,
reference images may not be available to users (e.g., at the receiver).
Second, most user-generated images do not have references.  The need for BIQA
grows rapidly due to the popularity of social media platforms and
multi-party video conferencing. 

Research on BIQA has received a lot of attention in recent years.
Existing BIQA methods can be categorized into two types: conventional
methods and deep-learning-based (DL-based) methods.  Most conventional
methods adopt a standard pipeline: a quality-aware feature extraction
followed by a regressor that maps from the feature space to the quality
score space. To give an example, methods based on natural scene
statistics (NSS) analyze statistical properties of distorted images and
compute the distortion degree as quality-aware features.  These
quality-aware features can be represented by discrete wavelet transform
(DWT) coefficients~\cite{moorthy2010two}, discrete cosine transform
(DCT) coefficients~\cite{saad2012blind}, luminance coefficients in the
spatial domain~\cite{mittal2012no}, and so on.  Codebook-based
methods~\cite{ye2012no, ye2012unsupervised, zhang2014training,
xu2016blind} generate features by extracting representative codewords
from distorted images. After that, a regressor is trained to project
from the feature domain to the quality score domain. 

Inspired by the success of deep neural networks (DNNs) in computer
vision, researchers have developed DL-based
methods to solve the BIQA problem. On one hand, the DL-based methods
achieve high performance because of their strong feature representation
capability and efficient regression fitting. On the other hand, existing
annotated IQA datasets may not have sufficient content to train large
DNN models. Given that collecting large-scale annotated IQA datasets are
expensive and time-consuming and that DL-based BIQA methods tend to
overfit the training data from IQA datasets of limited sizes, it is
critical to address the overfitting problem caused by small-scale
annotated IQA datasets. Effective DL-based solutions adopt a large
pre-trained model that was trained on other datasets, e.g.
ImageNet~\cite{deng2009imagenet}. 

The transferred prior information from a pre-trained model improves the
test performance.  Nevertheless, it is difficult to implement a large
pre-trained model of high complexity on mobile or edge devices.  As
social media contents are widely accessed via mobile terminals, it is
desired to conduct BIQA with limited model sizes and computational
complexity. A lightweight high-performance BIQA solution is in great
need.  To address this void, we study the BIQA problem in depth and
propose a new solution called ``GreenBIQA". This work has the following
three main contributions. 
\begin{itemize}
\item A novel GreenBIQA method is proposed for images with synthetic and
real-world (or called authentic) distortions. It offers a transparent
and modularized design with a feedforward training pipeline. The
pipeline includes unsupervised representation generation, supervised
feature selection, distortion-specific prediction, regression, and ensembles of
prediction scores. 
\item We conduct experiments on four IQA datasets to demonstrate the
prediction performance of GreenBIQA. It outperforms all conventional
BIQA methods and DL-based BIQA methods without pre-trained models in
prediction accuracy. As compared to state-of-the-art BIQA methods with
pre-trained networks, the prediction performance of GreenBIAQ is still
quite competitive yet demands a much smaller model size and
significantly lower inference complexity. 
\item We carry out experiments under the weakly-supervised learning
setting to demonstrate the robust performance of GreenBIQA as the number
of training samples decreases. Also, we show how to exploit active
learning in selecting images for labeling. 
\end{itemize} 

It is worthwhile to point out that preliminary results of our research
were presented in \cite{mei2022greenbiqa}. This work is its extension.
The additional content includes a more thorough literature review in
Sec.  \ref{G_sec:related}, an elaborative description of the GreenBIQA
method and more exemplary images to illustrate key discussed ideas in
Sec. \ref{G_sec:BIQA_method}, improved and extended experimental results
in Sec. \ref{G_sec:experiments}.  In particular, we have added new
experimental results on memory/latency tradeoff, cross-domain learning,
ablation study, weak-supervision, and active learning. 

The rest of this paper is organized as follows. Related work is reviewed
in Sec. \ref{G_sec:related}.  GreenBIQA is described in Sec.
\ref{G_sec:BIQA_method}.  Experimental results are shown in Sec.
\ref{G_sec:experiments}. Finally, concluding remarks are given in Sec.
\ref{G_sec:conclusion}.

\section{Related Work}\label{G_sec:related}

\subsection{Conventional BIQA Methods}

Conventional BIQA methods adopt a two-step processing pipeline: 1)
extracting quality-aware features from input images, and 2) using a
regression model to predict the quality score based on extracted
features. The support vector regressor (SVR)~\cite{awad2015support} or
the XGBoost regressor \cite{chen2016xgboost} is often employed in the
second step. According to the differences in the first step, we
categorize conventional BIQA methods into two main types. 

\subsubsection{Natural Scene Statistics (NSS)}

The first type relies on natural scene statistics (NSS). These methods
predict image quality by evaluating the distortion of the NSS
information.  For example, DIIVINE~\cite{moorthy2011blind} proposed a
two-stage framework, including a classifier to identify different
distortion types which is followed by a distortion-specific quality
assessment.  Instead of computing distortion-specific features,
NIQE~\cite{mittal2012making} evaluated the quality of distorted images
by computing the distance between the model statistics and those of
distorted images. BRISQUE~\cite{mittal2012no} used NSS to quantify the
loss of naturalness caused by distortions, which is operated in the
spatial domain with low complexity.  BLINDS-II~\cite{saad2012blind}
proposed an NSS model using the discrete cosine transform (DCT)
coefficients and, then, adopted the Bayesian inference approach to
predict image quality using features extracted from the model.
NBIQA~\cite{ou2019novel} developed a refined NSS mode by collecting
competitive features from existing NSS models in both spatial and
transform domains.  Histogram counting and the Weibull distribution were
employed in \cite{xue2014blind} and \cite{zhang2015feature},
respectively, to analyze the statistical information and build the
distribution models.  Although above-mentioned methods utilized the NSS
information in wide variety, they are still not powerful enough to
handle a broad range of distortion types, especially for datasets with
authentic distortions. 

\subsubsection{Codebook-based Methods}

The second type extracts representative codewords from distorted images.
The common framework of codebook-based methods includes: local feature
extraction, codebook construction, feature encoding, spatial pooling,
and quality regression.  CBIQ \cite{ye2012no} constructed visual
codebooks from training images by quantizing features, computed the
codeword histogram, and fed the histogram data to the regressor.
Following the same framework, CORNIA~\cite{ye2012unsupervised} extracted
image patches from unlabeled images as features, built a codebook (or a
dictionary) based on clustering, converted an image into a set of
non-linear features, and trained a linear support vector machine to map
the encoded quality-aware features to quality scores.  Non-linear
features in this pipeline were obtained from the dictionary using
soft-assignment coding with spatial pooling. However, the codebook needs
a large number of codewords to achieve good performance.  The high order
statistics aggregation (HOSA) was exploited in \cite{xu2016blind} to
design a codebook of a smaller size.  That is, besides the mean of each
cluster, the high-order statistical information (e.g., dimension-wise
variance and skewness) inside each cluster can be aggregated to reduce
the codebook size.  Generally speaking, codebook-based methods rely on
high-dimensional handcrafted feature vectors, and they are not effective
in handling diversified distortion types. 

\subsection{DL-based BIQA Methods}

DL-based methods have been intensively studied to solve the BIQA
problem. A solution based on the convolutional neural network (CNN) was
first proposed in \cite{kang2014convolutional}. It includes one
convolutional layer with max and min pooling and two fully connected
layers.  To alleviate accuracy discrepancy between FR-IQA and NR-IQA, a
local quality map was derived using CNN to imitate the behaviors of
FR-IQA in BIECON~\cite{kim2016fully}. Then, a statistical pooling
strategy is adopted to capture the holistic properties and generate
fixed-size feature vectors. A DNN model was proposed in
WaDIQaM~\cite{bosse2017deep} by including ten convolutional layers as well as
five pooling layers for feature extraction, and two fully connected
layers for regression. MEON~\cite{ma2017end} proposed two sub-networks
to achieve better performance on synthetic datasets.  The first
sub-network classifies the distortion types while the second sub-network
predicts the final quality.  By sharing their earlier layers, the two
sub-networks can solve their sub-tasks jointly for better performance. 

Quality assessment of images with authentic (i.e., real-world) distortions
is challenging due to mixed distortion types and high content variety.
Recent DL-based methods all adopt advanced DNNs.  Feature extraction
using a pre-trained ResNet~\cite{he2016deep} was adopted in
\cite{zeng2017probabilistic}.  A probabilistic quality representation
was proposed in PQR~\cite{zeng2018blind}, which employed a more robust
and optimal loss function to describe the score distribution generated
by different subjective.  It improved the quality prediction accuracy
and sped up the training process.  A self-adaptive hyper network
architecture was utilized by HyperIQA~\cite{su2020blindly} to adjust the
quality prediction parameters.  It can handle a broad range of
distortions with a local distortion-aware module and deal with wide
content variety with perceptual quality patterns based on recognized
content adaptively. DBCNN~\cite{zhang2018blind} adopted DNN models
pre-trained by large datasets to facilitate quality prediction on both
synthetic and authentic datasets. A network pre-trained by
synthetic-distortion datasets was used to classify distortion types and
levels.  Another pre-trained network based on the
ImageNet~\cite{deng2009imagenet} was used as the classifier.  The two
feature sets from two models were integrated into one representation for
final quality prediction through bilinearly pooling.  The absence of the
ground truth reference was compensated in
Hallucinated-IQA~\cite{lin2018hallucinated}, which generated a
hallucinated reference using generative adversarial networks
(GANs)~\cite{goodfellow2020generative}. 

Instead of predicting the mean opinion score (MOS) generated by
subjects, NIMA~\cite{talebi2018nima} predicted the MOS distribution
using a CNN.  To balance the trade-off between performance accuracy and
number of model parameters, NIMA had three models with different architectures,
namely, VGG16~\cite{simonyan2014very},
Inception-v2~\cite{szegedy2016rethinking}, and  MobileNet~\cite{howard2017mobilenets}. 
NIMA (VGG16) gave the best
performance but with the longest inference time and the largest model
size.  NIMA (MobileNet) was the
smallest one with the fewest model parameters but the worst accuracy.
Although NIMA (MobileNet) has a small model size, it is still difficult
to deploy it on mobile/edge devices. 

\begin{figure*}[!htbp]
\centering
\includegraphics[width=0.95\linewidth]{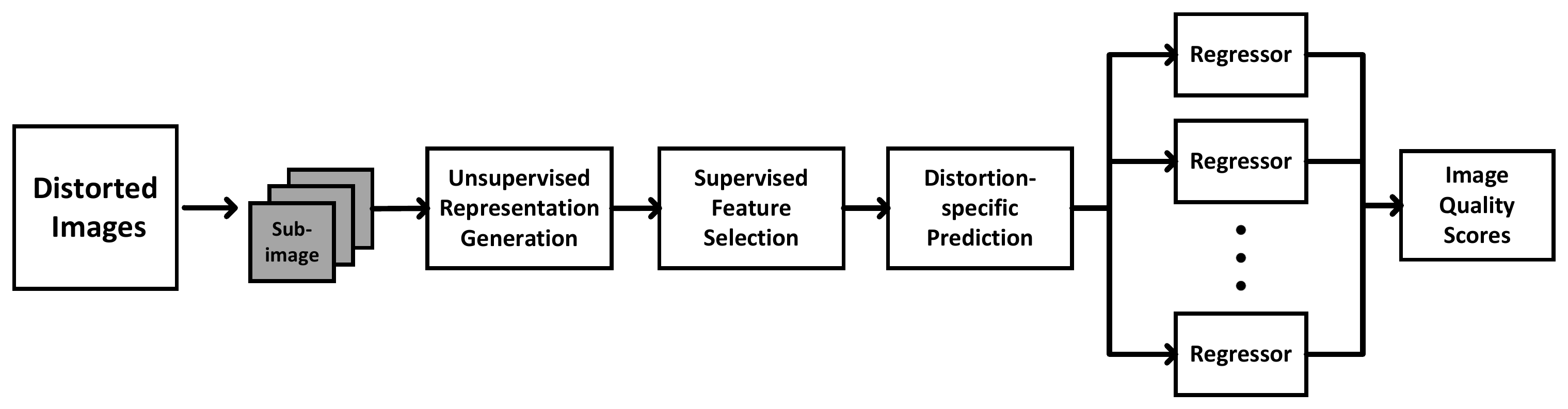}\\
\caption{An overview of the proposed GreenBIQA method.}\label{fig:pipeline}
\end{figure*}

\subsection{Green Machine Learning}

Green learning~\cite{kuo2023green} has been proposed recently as an
alternative machine learning paradigm that targets efficient models of
low carbon footprint. They are characterized by small model sizes and
low training and inference computational complexities. An additional
advantage is its mathematical transparency through a modularized design
principle. Green learning was originated by efforts in understanding the
functions of various components of CNNs such as nonlinear activation
\cite{kuo2016understanding}, convolutional layers and fully-connected
layers \cite{kuo2019interpretable}. Its development path started to
deviate from neural networks by giving up the basic neuron unit and the
network architecture since 2020. Examples of green learning models
include PixelHop \cite{chen2020pixelhop} and PixelHop++
\cite{chen2020pixelhop++} for object classification and PointHop
\cite{zhang2020pointhop} and PointHop++ \cite{zhang2020pointhop++} for
3D point cloud classification. Green learning techniques have been
developed for many applications such as deepfake detection
\cite{chen2021defakehop}, anomaly detection \cite{zhang2021anomalyhop},
image generation \cite{lei2021tghop}, etc. We propose a lightweight BIQA
method in this work by following this path.

\begin{figure}[!ht]
\centering
\includegraphics[width=0.6\linewidth]{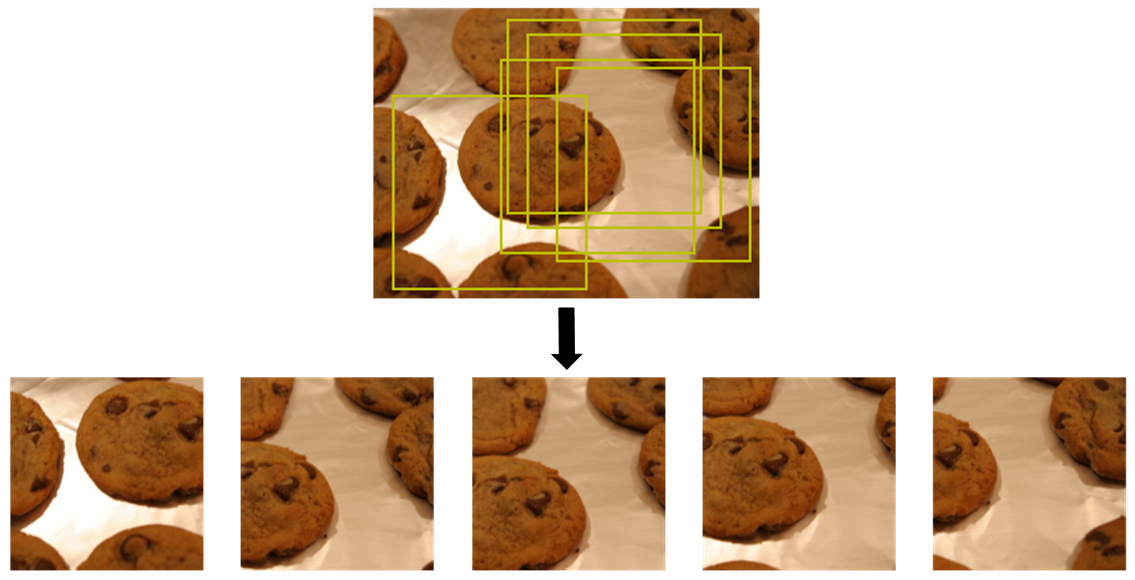}\\
\caption{An example of five cropped sub-images for authentic-distortion
datasets.}\label{fig:cropping_authentic}
\end{figure}

\begin{figure}[!ht]
\centering
\includegraphics[width=0.6\linewidth]{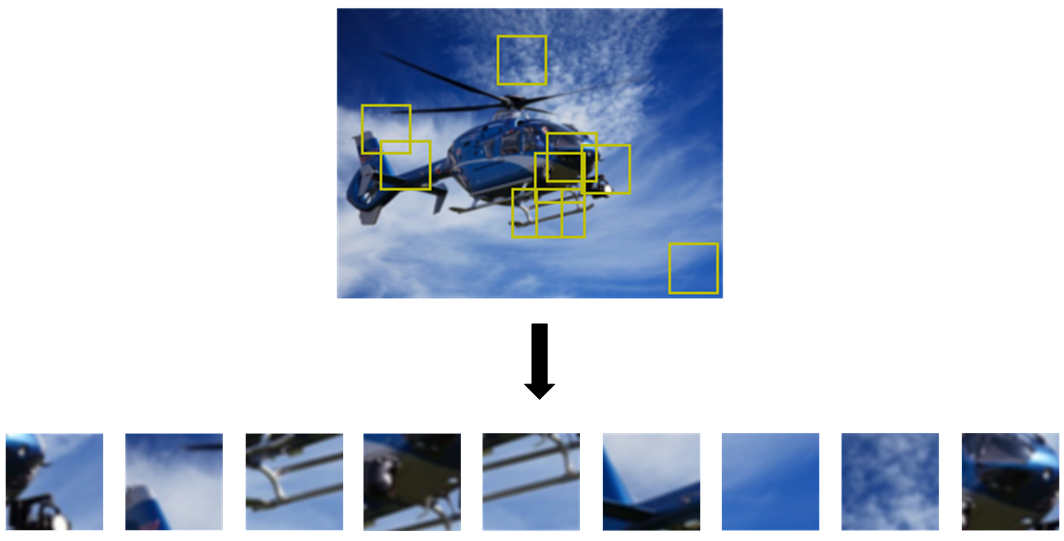}\\
\caption{An example of nine cropped sub-images for synthetic-distortion
datasets.}\label{fig:cropping_synthetic}
\end{figure}

\begin{figure*}[t]
\centering
  \begin{tabular}{c @{\hspace{20pt}} c }
    \includegraphics[width=.45\linewidth]{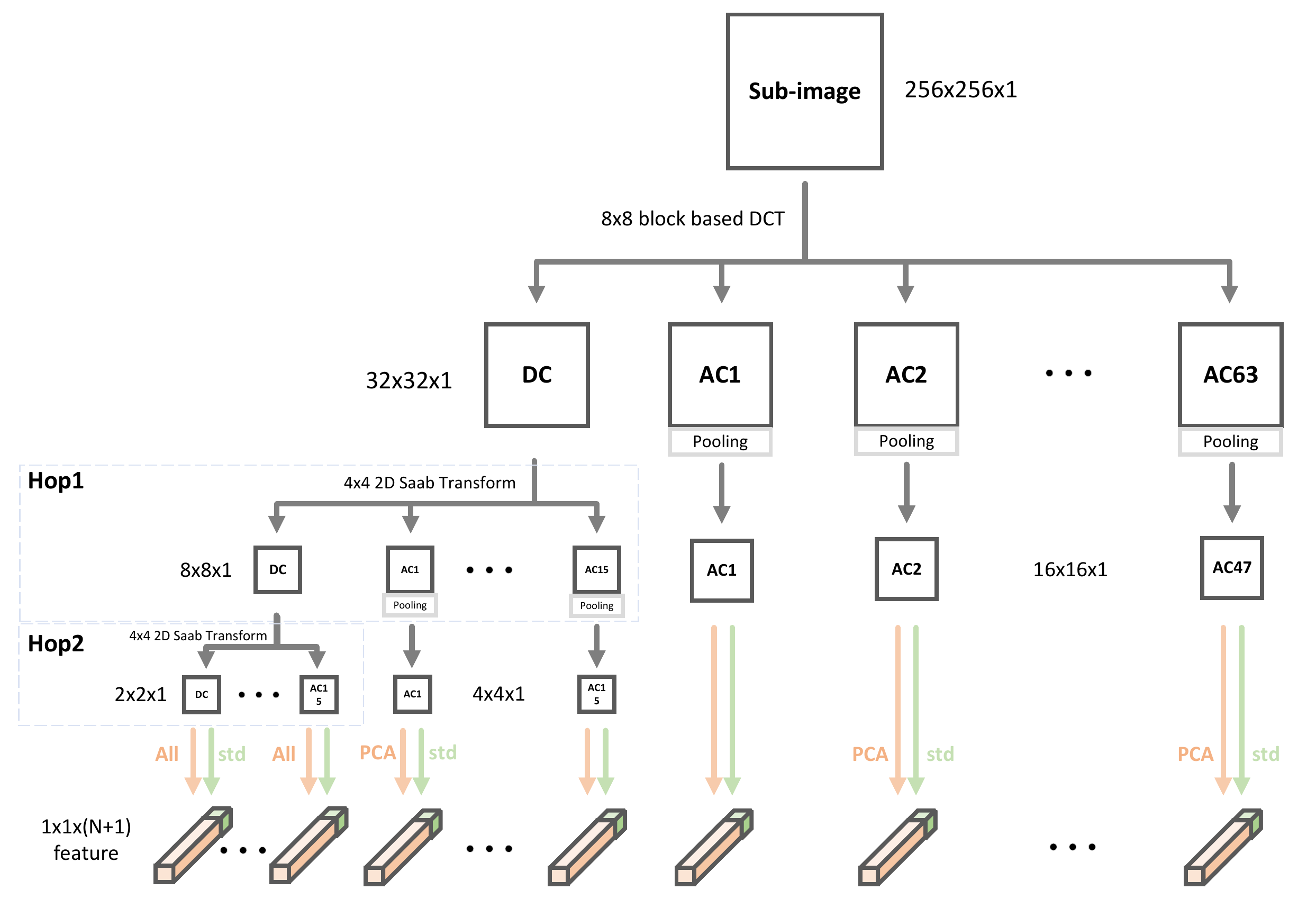} &
    \includegraphics[width=.45\linewidth]{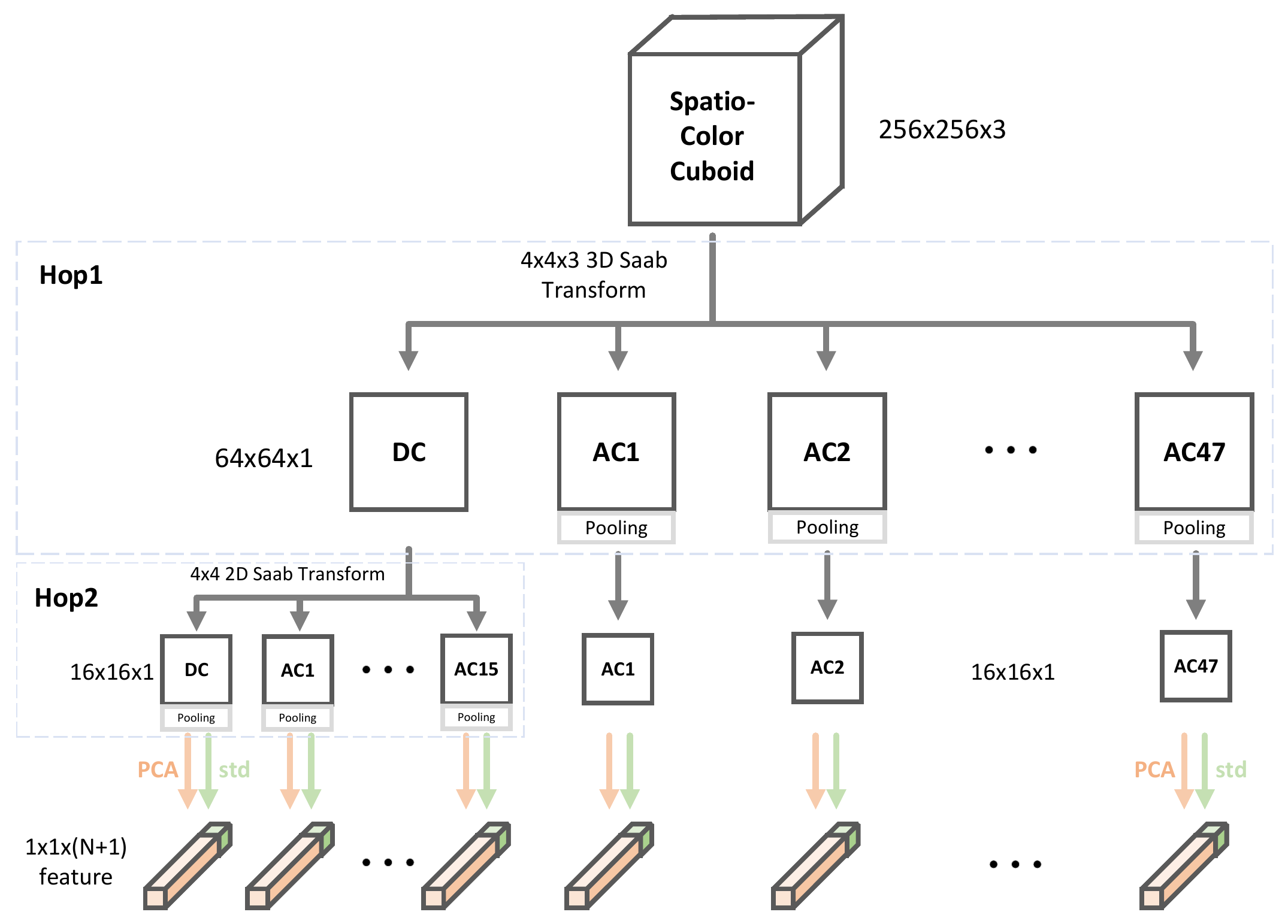}
    \\[\abovecaptionskip]
    \small (a) Generation of spatial representations &
    \small (b) Generation of joint spatio-color representations
  \end{tabular}
\caption{Unsupervised representation generation: spatial representations 
and joint spatio-color representations.}\label{fig:feature_extraction}
\end{figure*}

\section{Proposed GreenBIQA Method}\label{G_sec:BIQA_method}

An overview of the proposed GreenBIQA method is depicted in Fig.
\ref{fig:pipeline}. As shown in the figure, GreenBIQA has a modularized
solution that consists of five modules: (1) image cropping, (2)
unsupervised representation generation, (3) supervised feature
selection, (4) distortion-specific prediction, and (5) regression and
decision ensemble. They are elaborated below.

\subsection{Image Cropping}

Image cropping is implemented to standardize the input size and enlarge
the number of training samples.  It is achieved by cropping sub-images
of fixed size from raw images in datasets.  All cropped sub-images are
assigned the same mean opinion score (MOS) as their source image.  To
ensure the high correlation between sub-images and their assigned MOS,
we adopt different cropping strategies for synthetic-distortion and
authentic-distortion datasets as shown in Fig.
\ref{fig:cropping_authentic} and Fig. \ref{fig:cropping_synthetic},
respectively. 

For images in authentic-distortion datasets such as
KonIQ-10K~\cite{hosu2020koniq}, they contain distortions in unknown
regions.  Thus, we crop a smaller number of sub-images of a larger size
(e.g., $256 \times 256$ out of $384 \times 512$) to ensure the assigned
MOS for each sub-image is reasonable. The cropped sub-images can overlap
with one another. Fig.  \ref{fig:cropping_authentic} shows five randomly
cropped sub-images from one source image. 

For images in synthetic-distortion datasets such as
KADID-10K~\cite{lin2019kadid}, all distortions are applied to the
reference images uniformly with few exceptions (e.g., color distortion
in localized regions in KADID-10K).  Only one distortion type is added
to one image at a time. Therefore, cropping sub-images of a smaller size
is sufficient to capture distortion characteristics.  Furthermore, we
can crop more sub-images to enlarge the number of training samples and
conduct decision ensembles in the inference stage.  An example of image
cropping from the KADID-10K dataset is shown in Fig.
\ref{fig:cropping_synthetic}, where nine sub-images of size of $64
\times 64$ are randomly selected. 

\subsection{Unsupervised Representation Generation}\label{subsec:feature_generation}

Given sub-images from the image cropping module, we extract a set of
representations from sub-images in an unsupervised manner.  We consider
two types of representations.
\begin{enumerate}
\item Spatial representations. They are extracted from the Y, U, and V
channels of sub-images individually. 
\item Joint spatio-color representations. They are extracted from a 3D
cuboid of size $H \times W \times C$, where $H$ and $W$ are the height
and width of a sub-image and $C=3$ is the number of color channels,
respectively. 
\end{enumerate}

\subsubsection{Spatial Representations}\label{subsec:spatial_features}

Fig. \ref{fig:feature_extraction} (a) shows the procedure of spatial
representation generation. The representations are derived from $8 \times 8$
block DCT coefficients since they are often available in compressed
images. The input sub-images are first partitioned into non-overlapping
blocks of size $8 \times 8$ and DCT coefficients are generated by the
block DCT transform.  DCT coefficients of each block are scanned in the
zigzag order, leading to one DC coefficient and 63 AC coefficients,
denoted by AC1-AC63.  We split them into 64 channels. Generally, the amount
of energy decreases from the DC channel to the AC63 channel.  
There are correlations among DC coefficients of spatially adjacent
blocks.  We apply the Saab transform~\cite{kuo2019interpretable} to
them. The Saab transform uses a constant-element kernel to compute the
patch mean, which is called the DC component of the Saab transform.
Then, it applies the principal component analysis (PCA) to mean-removed
patches in deriving data-driven kernels, called AC kernels. The
application of AC kernels to each patch yields AC coefficients of the
Saab transform. Here, we decorrelate DC coefficients in two stages, i.e., 
Hop1 and Hop2. 
\begin{itemize}
\item Hop1 Processing: We partition $32 \times 32$ DC coefficients into
non-overlapping blocks of size $4 \times 4$ and conduct the Saab transform on
each block, leading to one DC channel and 15 AC channels in Hop1. We
feed the $8 \times 8$ DC coefficients to the next hop. 
\item HOP2 Processing: We apply another $4 \times 4$ Saab transform on each of
non-overlapping blocks of size $4 \times 4$, leading to DC and 15 AC channels in
Hop2.  We collect all the representations from Hop2 and append them to
the final representation set in preserving low-frequency details. 
\end{itemize}
Other Saab coefficients in Hop1 and other DCT coefficients at the top
layer contain mid- and high-frequency information. We need to aggregate
them spatially to reduce the representation number.  First, we take their
absolute values and apply the maximum pooling to lower their dimension
as indicated by the down-ward gray arrow. Next, we adopt the following 
operations to yield two sets of values:
\begin{itemize}
\item Compute the maximum value, the mean value, and the standard
deviation of the same coefficients across the spatial domain. 
\item Conduct the PCA transform on spatially adjacent regions for
further dimension reduction (except the coefficients in HOP2). 
\end{itemize}
These values are concatenated to form spatial representations of
interest.  The same process is applied to the Y, U, and V channels of
all sub-images.

\subsubsection{Joint Spatio-Color Representations}

We first convert sub-images from the YUV to RGB color space.  The
corresponding spatio-color cuboids have a size of $H \times W \times C$,
where $H$ and $W$ are the height and width of the sub-image,
respectively, and $C=3$ is the number of color channels.  They serve as
input cuboids to a two-hop hierarchical structure as shown in Fig.
\ref{fig:feature_extraction} (b). In Hop1, we split the input cuboids
into non-overlapping cuboids of size $4 \times 4 \times 3$ and apply the 3D Saab
transform to them individually - leading to one DC channel and 47 AC
channels, denoted by AC1-AC47. Each channel has a spatial dimension of
$64 \times 64$.  Since the DC coefficients are spatially correlated, we apply the
2D Saab transform in Hop2, where the DC channel of size $64 \times 64$ is
decomposed into $16 \times 16$ non-overlapping blocks of size $4 \times 4$.  For other 47
AC coefficients in the output of Hop1, we take their absolute values and
conduct the 4x4 max pooling, leading to 47 channels of spatial dimension
$16 \times 16$. In total, we obtain $16+47=63$ channels of the same spatial size 
$16 \times 16$. We use the following two steps to extract joint spatio-color
features.
\begin{itemize}
\item Flatten blocks to vectors, conduct PCA, and select coefficients
from the first $N$ principal components.
\item Compute the standard deviation of the coefficients in the same channel.
\end{itemize}
The above two sets of representations are concatenated to form the joint
spatio-color representations. 

\begin{figure}[!htbp]
\centering
\includegraphics[width=0.6\linewidth]{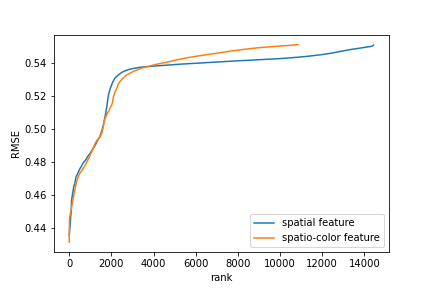}\\
\caption{RFT results of spatial and spatio-color representations.}
\label{fig:rft_3}
\end{figure}

\subsection{Supervised Feature Selection}\label{subsec:feature_selection}

It is desired to select more discriminant features from a large
number of representations obtained from the second module. A powerful
tool, called the relevant feature test (RFT)~\cite{yang2022supervised},
is adopted to achieve this objective. It computes the loss of
each representation independently. A lower loss value indicates a better
representation.  To conduct RFT, we split the dynamic range of a
representation into two sub-intervals with a set of partition points.
For a given partition, we first calculate the means of training samples in its
left and right regions respectively as the representative values and compute their
mean-squared errors (MSE) accordingly. Then, by combining the MSE values
of both regions together, we get the weighted MSE for the partition.
Then, we search for the smallest weighted MSE through a set of partition
points, and this minimum value defines the cost function of this
representation. Note that RFT is a supervised feature selection
algorithm since it exploits the label of the training samples.  We sort
representation indices according to their root-MSE (RMSE) values from
the smallest to the largest in Fig. \ref{fig:rft_3}. There are two
curves, one for the spatial representations and the other for the
spatio-color representations. We can use the elbow point on each curve
to select a subset of representations. In the experiment, we use RFT to
select 2048D spatial features and 2000D spatio-color features. The
former is a concatenation of spatial features from Y, U, and V channels.

\begin{figure*}[!ht]
\centering
\includegraphics[width=0.9\linewidth]{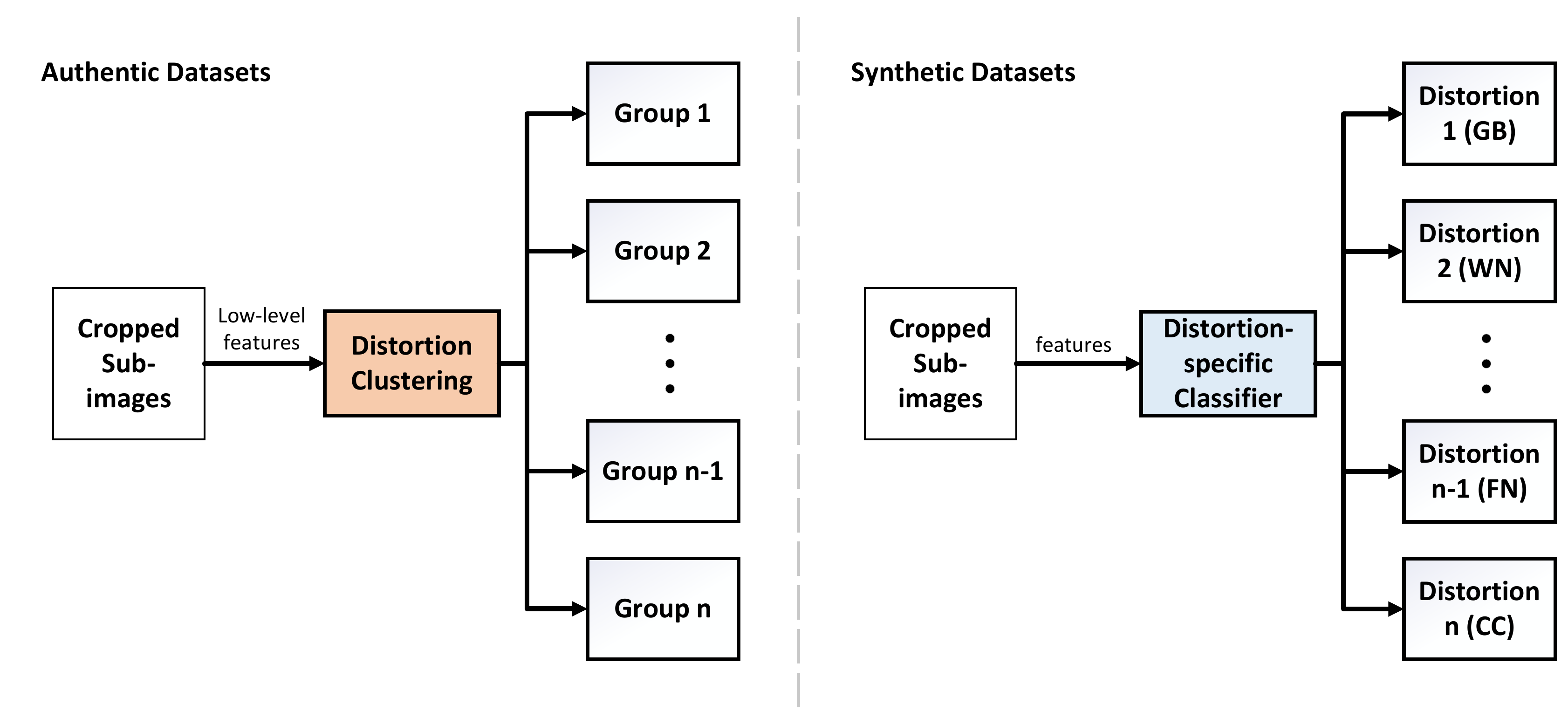}\\
\caption{Distortion-specific classifier and distortion clustering for
synthetic and authentic datasets, respectively.}\label{fig:distortion_classifier}
\end{figure*}

\begin{figure*}[ht]
\centering
\includegraphics[width=.25\linewidth]{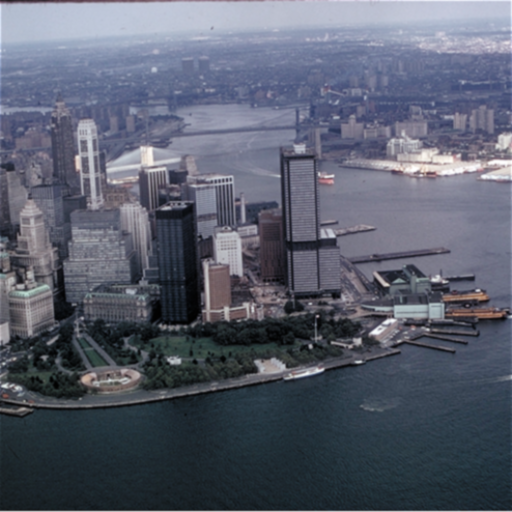} \hspace{5mm}
\includegraphics[width=.25\linewidth]{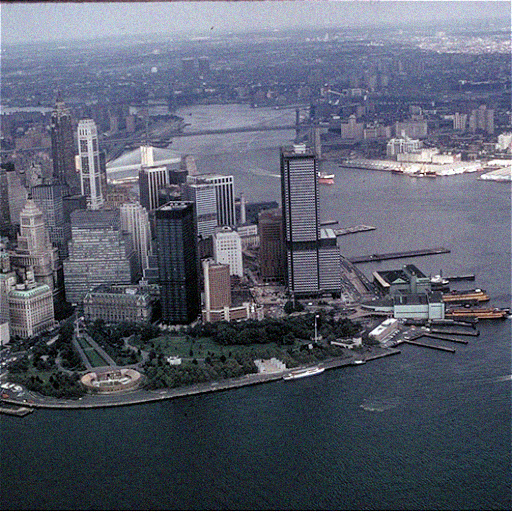} \hspace{5mm}
\includegraphics[width=.25\linewidth]{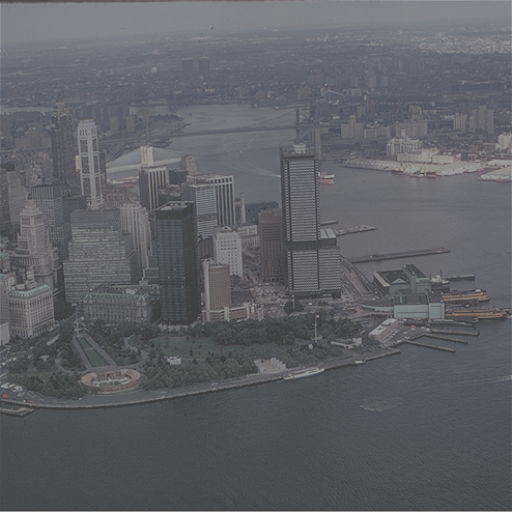} \\
(a) \hspace{4.5cm} (b) \hspace{4.5cm} (c) \\
\includegraphics[width=.25\linewidth]{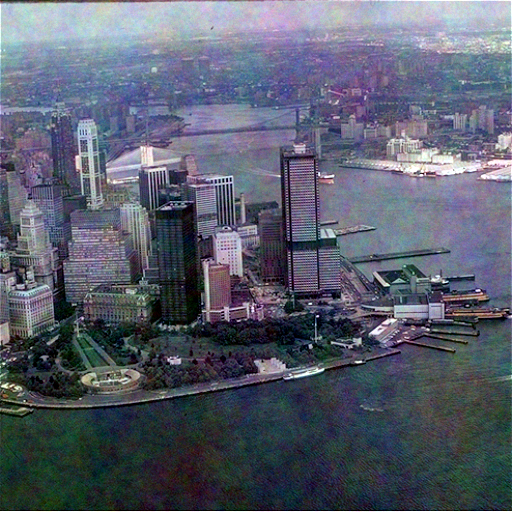} \hspace{5mm}
\includegraphics[width=.25\linewidth]{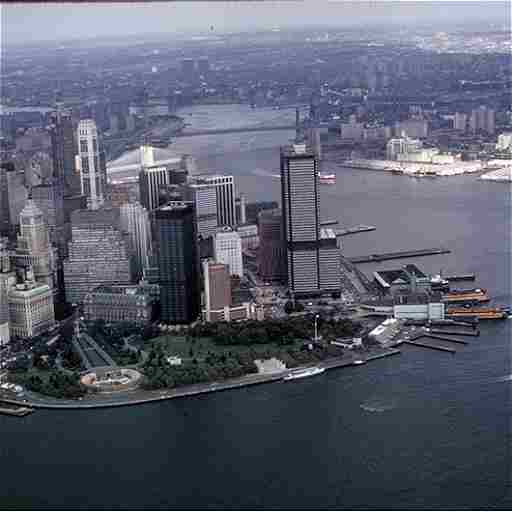} \hspace{5mm}
\includegraphics[width=.25\linewidth]{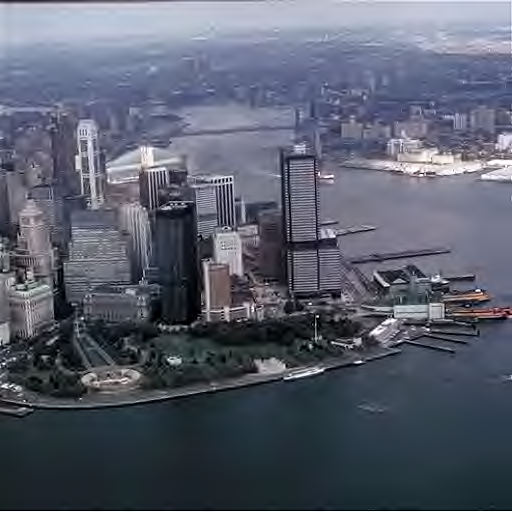} \\
(d) \hspace{4.5cm} (e) \hspace{4.5cm} (f) \\
\caption{Six synthetic distortion types in CSIQ: (a) Gaussian blur, (b) Gaussian noise, 
(c) Contrast decrements, (d) Pink Gaussian noise, (e) JPEG, and (f) JPEG-2000.}
\label{fig:synthetic_distortion}
\end{figure*}

\begin{figure*}[t]
\centering
\includegraphics[width=.30\linewidth]{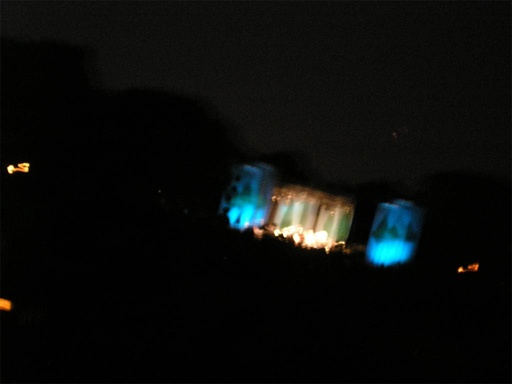} \hspace{5mm}
\includegraphics[width=.30\linewidth]{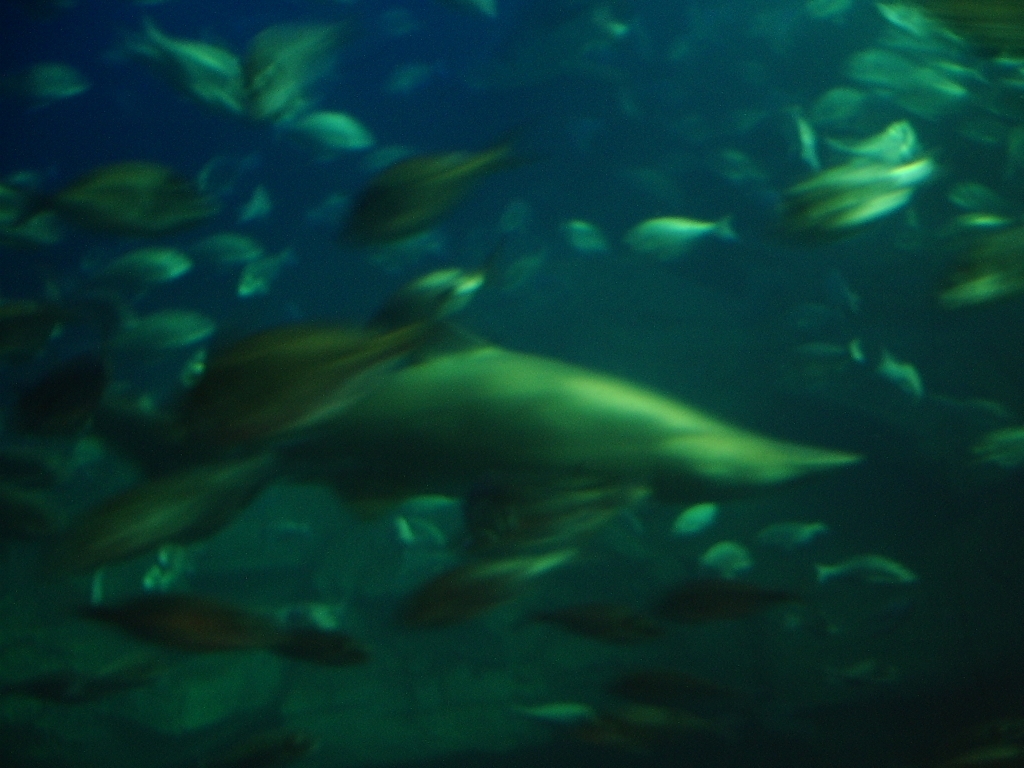} \hspace{5mm}
\includegraphics[width=.30\linewidth]{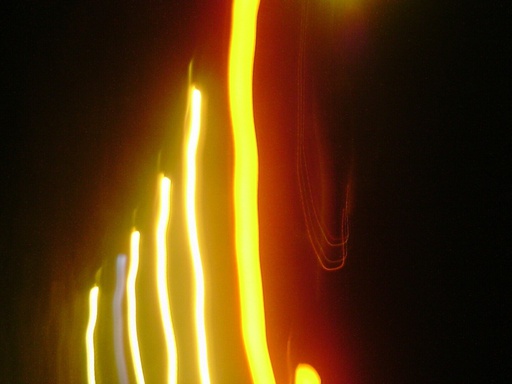} \\
(a) \hspace{5cm} (b) \hspace{5cm} (c) 
\caption{Three distorted images in KonIQ-10k: (a) Dark environment, (b) Underwater, 
and (c) Smeared light.}\label{fig:authentic_distortion}
\end{figure*}

\subsection{Distortion-specific Prediction}

Usually, we get better prediction scores if we can classify distorted
images into several classes based on their distortion types.  We examine
synthetic-distortion and authentic-distortion datasets separately as
shown in Fig. \ref{fig:distortion_classifier} due to their different
properties. 

\subsubsection{Synthetic Distortions} 

Images in synthetic-distortion datasets are usually associated with one
specific distortion type with multiple severity levels.  For example,
CSIQ~\cite{larson2010most} has 6 distortion types with 4 to 5 different
levels, as shown in Fig. \ref{fig:synthetic_distortion}. We can leverage
the known distortion types by first training a distortion classifier to
separate images accordingly.  Then, we design an individual pipeline to
handle each distortion type.  We can use distortion labels of training
images to train a multi-class distortion classifier based on the
selected features in Sec. \ref{subsec:feature_selection}.  There are
multiple sub-images from one image and each of them may have a different
predicted distortion type. We adopt majority voting to determine the
image-level distortion type.  Note that some distortion types are easily
confused with each other (e.g., JPEG and JPEG2000). We can simply merge
them into a single type. As a result, the class number can be reduced. 

\subsubsection{Authentic Distortions}

Images from authentic-distortion datasets may contain mixed distortion
types introduced in image capture or transmission.  Three distorted
images from KonIQ-10K are shown in Fig.  \ref{fig:authentic_distortion}.
It is difficult to define each as one specific type.  For example, the
underwater image contains blurriness, noise, and color distortion.
Thus, instead of training a specific-distortion classifier, we cluster
images into multiple groups using some low-level features in an
unsupervised manner (e.g., the K-means algorithm).  The low-level
features include statistical information in the spatial and color
domains.  For spatial features, we apply the Laplacian and Sobel edge
filters to all pixels in each sub-image, take their absolute values, and
compute the mean, variance, and maximum.  For color features, we compute
the variance of each color channel (such as Y, U, and V).  In addition,
higher-order statistics can also be collected as color features.  All
these extracted features are concatenated into a feature vector for
unsupervised clustering.  Although unsupervised clustering does not
assign a distortion type to a cluster, it reduces the content diversity
of sub-images in the same cluster. 



\subsection{Regression and Decision Ensemble}

For each of 6 distortions, 19 distortions, and 4 clusters for CSIQ,
KADID-10K, and authentic-distortion datasets, we train an XGBoost
regressor ~\cite{chen2016xgboost} that maps from the feature space to
the MOS score, respectively.  In the experiment, we set hyper-parameters
of the XGBoost regressor to the following: 1) the max depth of each tree
is 5, 2) the subsampling ratio is 0.6, 3) the maximum tree number is 2000,
and 4) the early stop is adopted. Given the predicted MOS scores of all
sub-image from the same source image, a median filter is applied to
generate the ultimate predicted MOS score of the input image. 



\begin{table}[!ht]
\centering
\caption{Four benchmarking IQA datasets, where the number of distorted
images, the number of reference images, the number of distortion types
and collection methods of each dataset are listed.}\label{table:dataset_BIQA}
\begin{tabular}{ l | cccc }
\hline
Datasets   & Dist. & Ref. & Dist. Types &Scenario \\ \hline
CSIQ       &866 &30 &6 &Synthetic\\
KADID-10K  &10,125 &81 &25 &Synthetic\\ 
LIVE-C     &1,169 &N/A &N/A &Authentic\\
KonIQ-10K  &10,073 &N/A &N/A &Authentic\\ \hline
\end{tabular}
\end{table}

\section{Experiments}\label{G_sec:experiments}

\subsection{Experimental Setup}\label{G_subsec:setup}

\subsubsection{Datasets}

We evaluate GreenBIQA on two synthetic IQA datasets and two authentic
IQA datasets. Their statistics are given in Table
\ref{table:dataset_BIQA}.  The two synthetic-distortion datasets are
CSIQ~\cite{larson2010most} and KADID-10K~\cite{lin2019kadid}. Multiple
distortions of various levels are applied to a set of reference images
to yield distorted images. CSIQ has six distortion types with four to
five distortion levels. KADID-10K contains 25 distortion types with five
levels for each distortion type.  LIVE-C~\cite{ghadiyaram2015massive}
and KonIQ-10K \cite{hosu2020koniq} are two authentic-distortion
datasets. They contain a broad range of distorted real-world images
captured by users. No reference image and specific distortion type are
available for each image. LIVE-C and KonIQ-10K have 1,169 and 10,073
distorted images, respectively. 

\subsubsection{Benchmarking Methods}

We compare the performance of GreenBIQA with nine benchmarking methods
in Table \ref{table:BIQA_individual}.  They include four conventional
and five DL-based BIAQ methods. We divide them into four categories.
\begin{itemize}
\item NIQE~\cite{mittal2012making} and BRISQUE~\cite{mittal2012no}.
They are conventional BIQA methods using NSS features.
\item CORNIA~\cite{ye2012unsupervised} and HOSA~\cite{xu2016blind}.
They are conventional BIQA methods using codebooks.
\item BIECON~\cite{kim2016fully} and WaDIQaM~\cite{bosse2017deep}.
They are DL-based BIQA methods without pre-trained models (or simple 
DL methods).
\item PQR~\cite{zeng2018blind}, DBCNN~\cite{zhang2018blind}, and
HyperIQA~\cite{su2020blindly}.
They are DL-based BIAQ methods with pre-trained models (or advanced 
DL methods).
\end{itemize}

\subsubsection{Evaluation Metrics}

The performance is measured by two popular metrics: the Pearson Linear
Correlation Coefficient (PLCC) and the Spearman Rank Order Correlation
Coefficient (SROCC).  PLCC evaluates the correlation between predicted
scores from an objective method and user's subjective scores (e.g., MOS) in
form of
\begin{equation}
\textit{PLCC} = 1 - \frac{\sum_{i}(p_i - p_m)(\hat{p_i}-\hat{p_m})}
{\sqrt{\sum_{i}(p_i - p_m)^2}\sqrt{\sum_{i}\hat{p_i}-\hat{p_m})^2}},
\end{equation}
where $p_i$ and $\hat{p_i}$ represent predicted and subjective scores
while $p_m$ and $\hat{p_m}$ are their means, respectively. 
SROCC measures the monotonicity between predicted scores
from an objective method and user's subjective scores via
\begin{equation}
\textit{SROCC} = 1 -  \frac{6\sum_{i=1}^{L}(m_i - n_i)^2} {L(L^2-1)},
\end{equation}
where $m_i$ and $n_i$ denote the ranks of the prediction and the ground
truth label, respectively, and $L$ denotes the total number of
samples or the number of images in our current case. 

\begin{figure}[!htbp]
\centering
\includegraphics[width=0.9\linewidth]{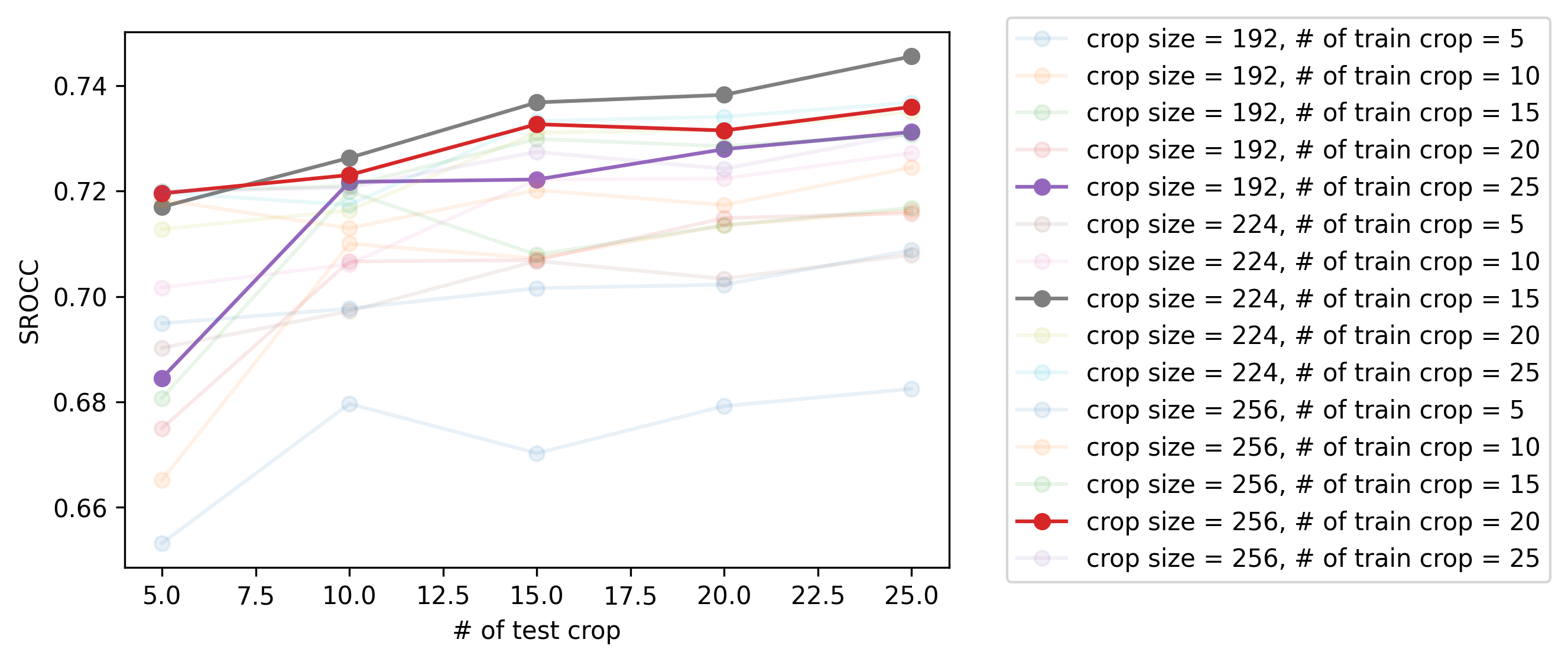}\\
\caption{Performance curve on the validation dataset of LIVE-C with 
different crop numbers and sizes.}\label{fig:crop_size}
\end{figure}

\subsubsection{Implementation Details}

In the training stage, we crop 15 sub-images of size $224 \times 224$
for each image in the two authentic datasets. This design choice is
based on SROCC performance of validation sets as shown in Fig.
\ref{fig:crop_size}, where the best performance under different crop
sizes is highlighted.  Similarly, we crop 25 sub-images of size $32
\times 32$ for each image in the two synthetic datasets. In the testing
(or inference) stage, we crop 25 sub-images of size $224 \times 224$ and
$32 \times 32$ for images in authentic and synthetic datasets,
respectively. 

We adopt the standard evaluation procedure by splitting each dataset
into 80\% for training and 20\% for testing.  Furthermore, 10\% of
training data is used for validation. We run experiments 10 times and
report median PLCC and SROCC values. For synthetic-distortion datasets,
splitting is implemented on reference images to avoid content overlap. 

\subsection{Performance Evalution}

We compare the performance of GreenBIQA and nine benchmarking BIQA
methods on four IQA datasets in Table \ref{table:BIQA_individual}.

\subsubsection{Comparison among Benchmarking Methods} We first compare
the performance among the nine benchmarks.  Although some conventional
BIAQ methods have comparable performance with simple DL methods (without
pre-trained models), we see a clear performance gap between conventional
BIQA methods and advanced DL methods (with pre-trained models). On the
other hand, the model size of advanced DL methods is significantly
larger. We comment on the performance of GreenBIQA against other
benchmarking methods below. 

\subsubsection{Synthetic-Distortion Datasets} For the two
synthetic-distortion datasets, CSIQ and KADID-10K, GreenBIQA achieves
the best performance among all.  This is attributed to its two
characteristics: 1) classification of synthetic distortions to multiple
types followed by different processing pipelines, and 2) effective usage
of ensemble decisions. For the first point, there are six distortion
types in CSIQ as shown in Fig.  \ref{fig:synthetic_distortion}.  We show
the SROCC performance of the best BIQA method in each of four categories
against each of six distortion types in the CSIQ dataset in Table
\ref{table:distortion}.  GreenBIAQ outperforms all others in four
distortion types. It performs especially well for JPEG distortion
because it adopts the DCT spatial features which match the underlying
compression distortion well.  GreenBIQA is also effective against white
Gaussian noise (WN), pink Gaussian noise (FN), and contrast decrements
(CC) through the use of joint spatial and spatio-color features.
GreenBIQA still works well for Gaussian blur (GB) although no blur
detector is employed. For the second point, since the number of
reference images is limited and the distortion is uniformly spread out
across the whole image, ensemble decision works well in such a setting. 

\begin{table*}[!htbp]
\centering
\caption{Performance comparison in PLCC and SROCC metrics between our
GreenBIQA method and nine benchmarking methods on four IQA databases,
where the nine benchmarking methods are categorized into four groups as
discussed in Sec. \ref{G_subsec:setup} and the best performance numbers
are shown in boldface.}
\label{table:BIQA_individual}
\resizebox{\textwidth}{!}{
\begin{tabular}{l c | c c c c c c c c c c }\hline
& & \multicolumn{2}{c}{CSIQ}
& \multicolumn{2}{c}{LIVE-C} & \multicolumn{2}{c}{KADID-10K} &
\multicolumn{2}{c}{KonIQ-10K}\\\hline
&Model & SROCC & PLCC & SROCC & PLCC & SROCC & PLCC & SROCC & PLCC & Model size (MB)\\\hline
&NIQE     &0.627 &0.712 &0.455 &0.483 &0.374 &0.428 &0.531 &0.538 &-\\
&BRISQUE  &0.746 &0.829 &0.608 &0.629 &0.528 &0.567 &0.665 &0.681 &-\\ \hline
&CORNIA   &0.678 &0.776 &0.632 &0.661 &0.516 &0.558 &0.780 &0.795 &7.4\\
&HOSA     &0.741 &0.823 &0.661 &0.675 &0.618 &0.653 &0.805 &0.813 &0.23\\ \hline
&BIECON   &0.815 &0.823 &0.595 &0.613 &-     &-     &0.618 &0.651 &35.2\\
&WaDIQaM  &0.844 &0.852 &0.671 &0.680 &-     &-     &0.797 &0.805 &25.2\\ \hline
&PQR      &0.872 &0.901 &0.857 &\textbf{0.882} &-     &-     &0.880 &0.884 &235.9\\
&DBCNN    &0.946 &\textbf{0.959} &0.851 &0.869 &0.851 &0.856 &0.875 &0.884 &54.6\\
&HyperIQA &0.923 &0.942 &\textbf{0.859} &\textbf{0.882} &0.852 &0.845 &\textbf{0.906} &\textbf{0.917} &104.7\\\hline
&GreenBIQA (Ours) &\textbf{0.952} &\textbf{0.959} &0.801 &0.809 &\textbf{0.886} &\textbf{0.893} &0.858 &0.870 &1.82\\ 
\hline
\end{tabular}}
\end{table*}

\begin{table*}[!htbp]
\centering
\caption{Comparison of the SROCC performance for each of six individual
distortion types in the CSIQ dataset, where WN, JPEG, JP2K, FN, GB, and
CC denote white Gaussian noise, JPEG compression, JPEG-2000 compression,
pink Gaussian noise, Gaussian blur, contrast decrements, respectively. The 
last column shows the weighted average of the SROCC metrics.}\label{table:distortion}
\begin{tabular}{ l | cc cc cc | c} \hline
         & WN & JPEG & JP2K & FN & GB & CC  & Average \\ \hline
BRISQUE  &0.723 &0.806 &0.840 &0.378 &0.820 &0.804 &0.728 \\ 
HOSA     &0.604 &0.733 &0.818 &0.500 &0.841 &0.716 &0.702 \\
BIECON   &0.902 &0.942 &0.954 &0.884 &\textbf{0.946} &0.523 &0.858 \\ 
HyperIQA &0.927 &0.934 &0.960 &0.931 &0.915 &\textbf{0.874} &0.923 \\ \hline
GreenBIQA (Ours)     &\textbf{0.943} &\textbf{0.980} &\textbf{0.969} 
         &\textbf{0.965} &0.894 &0.857 &\textbf{0.934} \\ \hline
\end{tabular}
\end{table*}

\subsubsection{Authentic-Distortion Datasets} For the two
authentic-distortion datasets, LIVE-C and KonIQ-10K, GreenBIQA
outperforms conventional BIQA methods and simple DL methods.  This
demonstrates the effectiveness of its extracted quality-aware features
and decision pipeline in handling diversified distortions and contents.
There is however a performance gap between GreenBIQA and advanced DL
methods with pre-trained models. The authentic-distortion datasets are
more challenging because of non-uniform distortions across images and a
wide variety of content without duplication. Since pre-trained models
are trained by a much larger image database, they have advantages in
extracting features for non-uniform distortions and unseen contents.
Yet, they demand much larger model sizes as a tradeoff. 

\subsection{Cross-Domain Learning}

To evaluate transferability of BIQA methods, we train models on one
dataset and test them on another dataset.  Due to the huge differences
in synthetic-distortion and authentic-distortion datasets, we focus on
authentic-distortion datasets and conduct experiments on LIVE-C and
KonIQ-10K only.  We consider two experimental settings: I) trained with
LIVE-C and tested on KonIQ-10K, and II) trained with KonIQ-10K and
tested on LIVE-C.  The SROCC performance of GreenBIQA and five
benchmarking methods under the two settings are compared in Table
\ref{table:BIQA_cross}, where benchmarks include the three best BIQA
methods in Table \ref{table:BIQA_individual} (i.e. PQR, DBCNN, and
HyperIQA) and two conventional BIQA methods (i.e.  BRISQUE and HOSA).
By comparing the performance numbers in Tables
\ref{table:BIQA_individual} and \ref{table:BIQA_cross}, we see a
performance drop in the cross-domain condition for all methods. We see
that GreenBIQA has a performance gap of 0.019 and 0.053 against the best
one, HyperIQA, for Experimental Settings I and II, respectively. As
shown in Table \ref{table:dataset_BIQA}, KonIQ-10K is much larger than
LIVE-C. Experimental Setting I provides a more proper environment to
demonstrate the robustness (or generalizability) of a learning model.
We compare the performance gaps in Table \ref{table:BIQA_cross} under
Setting I with those in the KonIQ-10K/SROCC column in Table
\ref{table:BIQA_individual}. The gaps between PQR, DBCNN, HyperIQA, and
GreenBIQA narrow down from 0.022, 0.017, and 0.048 to 0.004, 0.001, and
0.019, respectively. We see a greater potential of GreenBIQA along this
direction. 

\begin{table}[!htbp]
\centering
\caption{Comparison of the SROCC performance under the cross-domain learning 
scenario.}\label{table:BIQA_cross}
\begin{tabular}{l c| c c} \hline
& Settings      & I         & II \\\hline
& Train Dataset & LIVE-C    & KonIQ-10K\\
& Test Dataset  & KonIQ-10K & LIVE-C\\\hline
&BRISQUE  &0.425 &0.526 \\ 
&HOSA     &0.651 &0.648 \\ 
&PQR      &0.757 &0.770 \\
&DBCNN    &0.754 &0.755\\
&HyperIQA &\textbf{0.772} &\textbf{0.785}\\\hline
&GreenBIQA(Ours) &0.753 &0.732 \\\hline
\end{tabular}
\end{table}

\begin{figure*}[!htbp]
\centering
\includegraphics[width=0.9\linewidth]{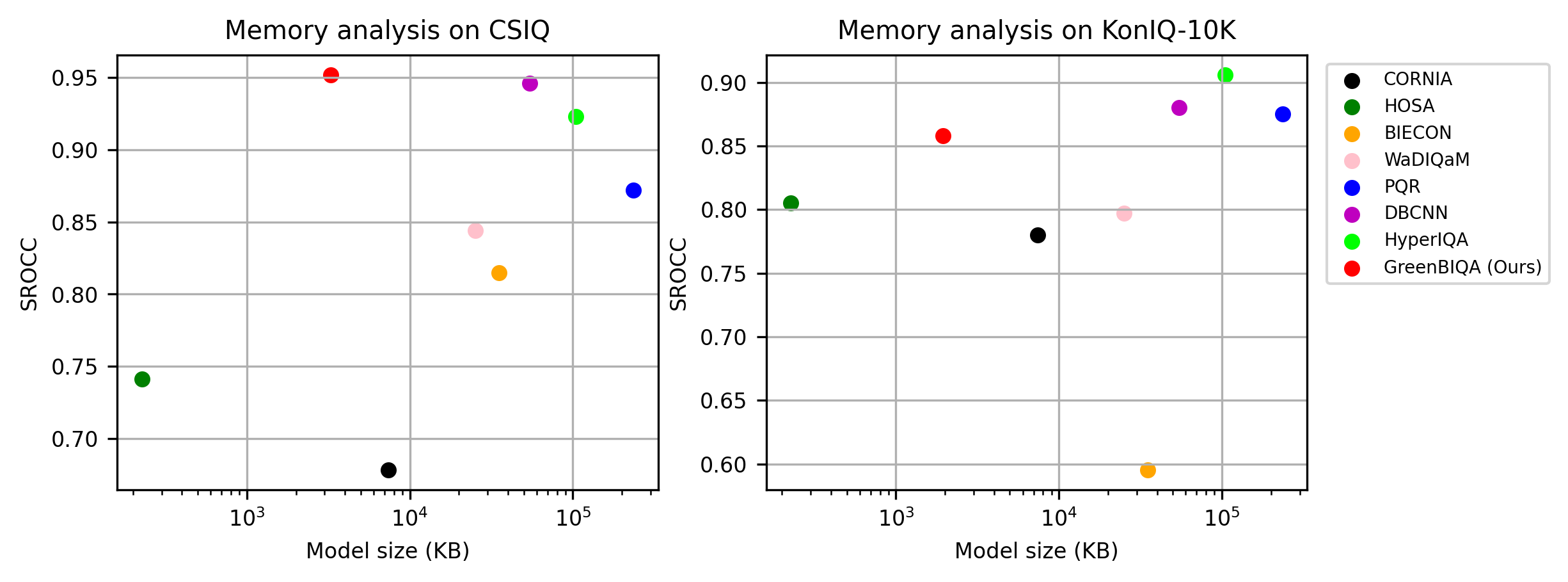}\\
(a) The SROCC performance versus the model size \\
\includegraphics[width=0.9\linewidth]{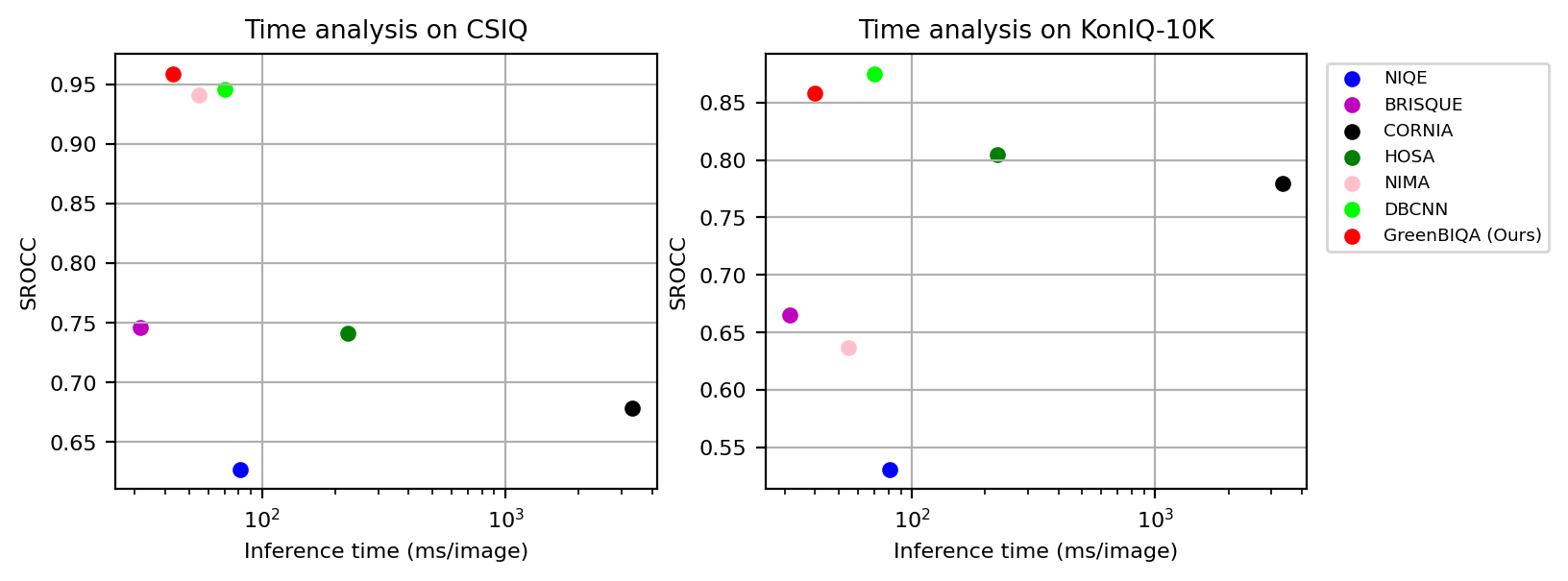}\\ 
(b) The SROCC performance versus the running time 
\caption{Illustration of the tradeoff between (a) the SROCC performance and
model sizes and (b) the SROCC performance and running time with respect
to CSIQ and KonIQ-10K datasets among several BIQA methods.}\label{fig:analysis}
\end{figure*}




\subsection{Model Complexity}

A lightweight model is critical to applications on mobile and edge
devices.  We analyze the model complexity of BIQA methods in four
aspects below: model sizes, inference time, computational complexity in
terms of floating-point operations (FLOPs), and memory/latency tradeoff. 

\subsubsection{Model Size}

There are two ways to measure the size of a learning model: 1) the
number of model parameters and 2) the actual memory usage.
Floating-point and integer model parameters are typically represented by
4 bytes and 2 bytes, respectively. Since a great majority of model
parameters are in floating point, the actual memory usage is roughly
equal to $4 \times \mbox{(no. of model parameters)}$ bytes (see Table
\ref{table:flop}). To avoid confusion, we use the ``model size" to refer
to actual memory usage below.  Fig. \ref{fig:analysis}(a) plots the
SROCC performance (in linear scale along the vertical axis) versus model
sizes (in log scale along the horizontal axis) on a synthetic-distortion
dataset (i.e., CSIQ) and an authentic-distortion dataset, (i.e.,
KonIQ-10K) with respect to a few benchmarking BIQA methods. The size of
the GreenBIQA model includes: the feature extractor (600KB), the
distortion-specific classifier (50KB) and several regressors (1.17MB),
leading to a total of 1.82 MB.  As compared with the two conventional
methods (CORINA and HOSA), GreenBIQA achieves much better performance
with comparable model sizes. GreenBIQA outperforms two simple DL methods
(BIECON and WaDIQaM), with a smaller model size.  As compared with the
three advanced DL methods (PQR, DBCNN and HyperIQA), GreenBIQA achieves
the best performance on CSIQ and competitive performance on KonIQ-10k at
a significantly smaller model size. Note that advanced DL methods have a
huge pre-trained network of size larger than 100MB as their backbones. 

\subsubsection{Inference Time}

Another important factor to consider is running time in inference, which
is especially the case for mobile/edge clients.  Fig.
\ref{fig:analysis}(b) shows the SROCC performance versus the inference
time (measured in milliseconds per image) for several benchmarking
methods on CSIQ and KonIQ-10K.  All methods are tested in the same
environment with a single CPU.  We compare GreenBIQA with four
conventional methods (NIQE, BRISQUE, CORNIA, and HOSA) and two DL methods
(NIMA and DBCNN).  GreenBIQA has clear advantages over all benchmarking
methods by considering the performance and inference time jointly.  It
is worthwhile to point out that GreenBIQA can process around 43 images
per second with a single CPU. In other words, it can meet the real-time
requirement by processing videos of 30 fps on a frame-by-frame basis.
The inference time of GreenBIQA can be further reduced by code
optimization and/or with the support of mature packages. 

\begin{table*}[t]
\centering
\caption{Comparison of SROCC/PLCC performance, no. of model parameters,
model sizes (memory usage), no. of GigaFlops and no. of KiloFlops per
pixel of several BIQA methods tested on the LIVE-C dataset, where ``X"
denotes the multiple no.}
\label{table:flop}
\resizebox{\textwidth}{!}{
\begin{tabular}{ l c | c c c c c c} \hline
&Model &SROCC &PLCC & Model Parameters (M) &Model Size (MB) &GFLOPs &KFLOPs/pixel\\ \hline
&NIMA(Inception-v2) &0.637 &0.698 &10.16 (22.6X) &37.4 (20.5X) &4.37 (128.5X) &87.10 (128.5X)\\
&BIECON &0.595 &0.613 &7.03 (15.6X) &35.2 (19.3X) &0.088 (2.6X) &85.94 (126.8X) \\
&WaDIQaM &0.671 &0.680 &5.2 (11.6X) &25.2 (13.8X) &0.137 (4X)   &133.82 (197.4X) \\
&DBCNN &0.851 &0.869 &14.6 (32.4X)  &54.6 (30X)   &16.5 (485.3) &328.84 (485.1X) \\
&HyerIQA &\textbf{0.859} &\textbf{0.882} &28.3 (62.9X) &104.7 (57.5X) &12.8 (376.5X) &255.10 (376.3X) \\ \hline
&GreenBIQA (Ours) &0.801 &0.809 &\textbf{0.45}(1X) & \textbf{1.82}(1X) 
         &\textbf{0.034} (1X) &\textbf{0.678}(1X) \\ \hline
\end{tabular}}
\end{table*}

\subsubsection{Computational Complexity}

We compare the SROCC and PLCC performance, the numbers of model parameters,
model sizes (in terms of memory usage), the numbers of Flops and Flops
per pixel of several BIQA methods tested on the LIVE-C dataset in Table
\ref{table:flop}. FLOPs is a common metric to measure the computational
complexity of a model. For a given hardware configuration, the number of
FLOPs is linearly proportional to energy consumption or carbon
footprint.  Column ``GFLOPs" in Table \ref{table:flop} gives the number
of GFLOPs needed to run a model once without considering the patch
number and size used in a method. For a fair comparison of FLOPs, we
compute the number of FLOPS per pixel defined by
\begin{equation}
FLOPs/pixel = \frac{FLOPs/patch}{H \times W},
\end{equation}
where $H$ and $W$ are the height and width of an input patch to a model,
respectively. NIMA with the pre-trained Inception-v2 network has low
performance, large model size, and high complexity.  Although simple DL
methods (e.g., WaDIQaM and BIECON) use smaller networks with lower
FLOPs, their performance is still inferior to GreenBIQA. Finally,
advanced DL methods (e.g., DBCNN and HyperIQA) outperform GreenBIQA in
SROCC and PLCC performance. However, their model sizes are much larger
and their computational complexities are much higher. The numbers of
FLOPs of DBCNN and HyperIQA are 485 and 376 multiples of that of
GreenBIQA, respectively. 

\begin{figure}[!htbp]
\centering
\includegraphics[width=0.6\linewidth]{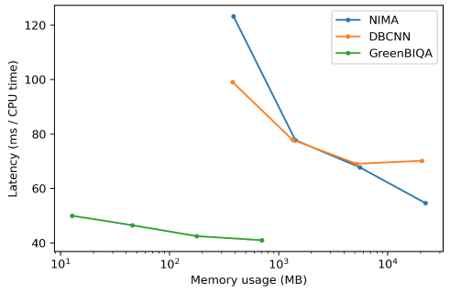}\\
\caption{Tradeoff between memory usage and latency for three BIQA
methods (NIMA, DBCNN and GreenBIQA), where latency can be reduced by the
usage of a larger memory.}\label{fig:mem_lat}
\end{figure}

\subsubsection{Memory/Latency Tradeoff}

There is a tradeoff between memory usage and latency in the image
quality inference stage. That is, latency can be reduced when given more
computing resources.  To observe the tradeoff, we control the memory
usage using different test image numbers in each run (i.e. the batch
size).  Fig. \ref{fig:mem_lat} shows the latency (in linear scale along
the vertical axis) and memory usage (in log scale along the horizontal
axis) of GreenBIQA and two advanced DL methods, where we set the batch
size equal to 1, 4, 16, and 64 in four experiments.  We see from the
figure that the latency of GreenBIQA is much smaller than NIMA and DBCNN
under the same memory size (say, $10^3$MB). Along this line, the memory
requirement of GreenBIQA is much lower than that of NIMA and DBCNN at
the same level of latency.  Again, the memory/latency tradeoff curve
of GreenBIQA can be further improved through code optimization.

\begin{table*}[!htbp]
\centering
\caption{Ablation Study for GreenBIQA.}
\resizebox{\textwidth}{!}{
\begin{tabular}{l c|c c c c c c c c}
\hline
& & \multicolumn{2}{c}{CSIQ}
& \multicolumn{2}{c}{LIVE-C}
& \multicolumn{2}{c}{KADID-1K} & \multicolumn{2}{c}{KonIQ-10k} \\\hline
&Components & SROCC & PLCC & SROCC & PLCC & SROCC & PLCC & SROCC & PLCC\\\hline
&S-features &0.925 &0.936 &0.774 &0.778 &0.847 &0.848 &0.822 &0.838\\
&S-features + SC-features &- &- &0.782 &0.783 &- &- &0.835 &0.850 \\ 
&S-features + Dist-predict &0.952 &0.959 &0.786 &0.788 &0.886 &0.893 &0.839 &0.856\\
&S-features + SC-features + Dist-predict &- &- &0.801 &0.809 &- &- &0.858 &0.870 \\\hline
\end{tabular}}
\label{table:BIQA_ablation}
\end{table*}

\subsection{Ablation Study}

To understand the impact of individual components on the overall
performance of GreenBIQA, we conduct an ablation study in Table
\ref{table:BIQA_ablation}, where S-features, SC-features, and
Dist-predict denotes spatial features, spatio-color features, and
distortion-specific prediction, respectively.  We first examine the
effectiveness of the spatial features and then add spatio-color features
in the first two rows. Both SROCC and PLCC improve on the two
authentic-distortion datasets.  Similarly, adding distortion-specific
prediction to S-features can improve SROCC and PLCC for all datasets in
the third row.  Finally, we use all the components in the fourth row and
see that SROCC and PLCC can be further improved to reach the highest
value.  Note that we do not report the performance of joint spatial and
spatio-color features for synthetic datasets since spatial features are
powerful enough.  The distortion-specific prediction benefits the
performance significantly on synthetic datasets by leveraging the
distortion label.

\begin{figure}[!htbp]
\centering
\includegraphics[width=0.6\linewidth]{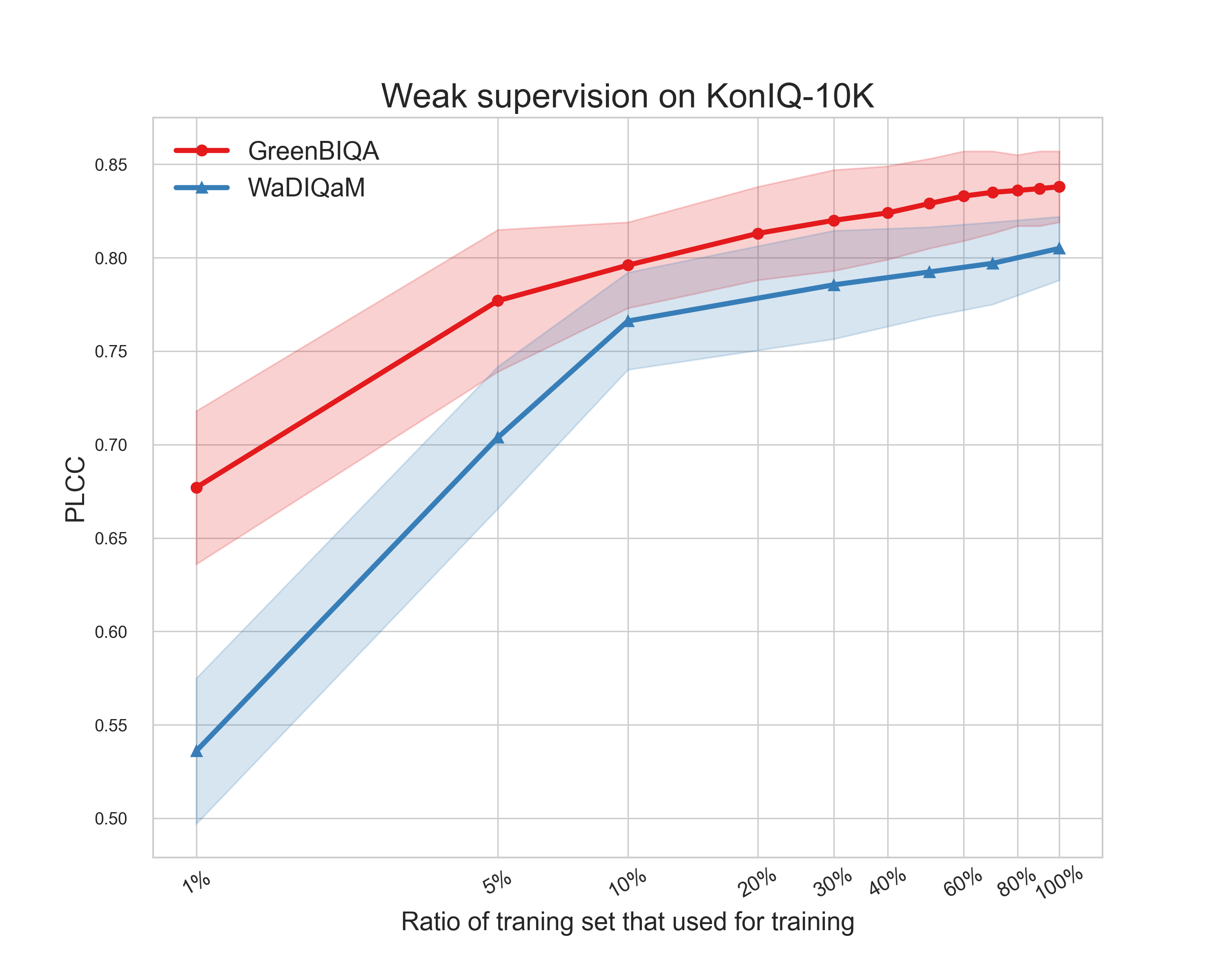}\\
\caption{The PLCC performance curves of GreenBIQA and WaDIQaM are
plotted as functions of the percentages of the full training
dataset of KonIQ-10K, where the solid line and the banded
structure indicate the mean value and the range of mean plus/minus 
one standard deviation, respectively.} \label{fig:weak_learning}
\end{figure}

\begin{figure}[!htbp]
\centering
\includegraphics[width=0.6\linewidth]{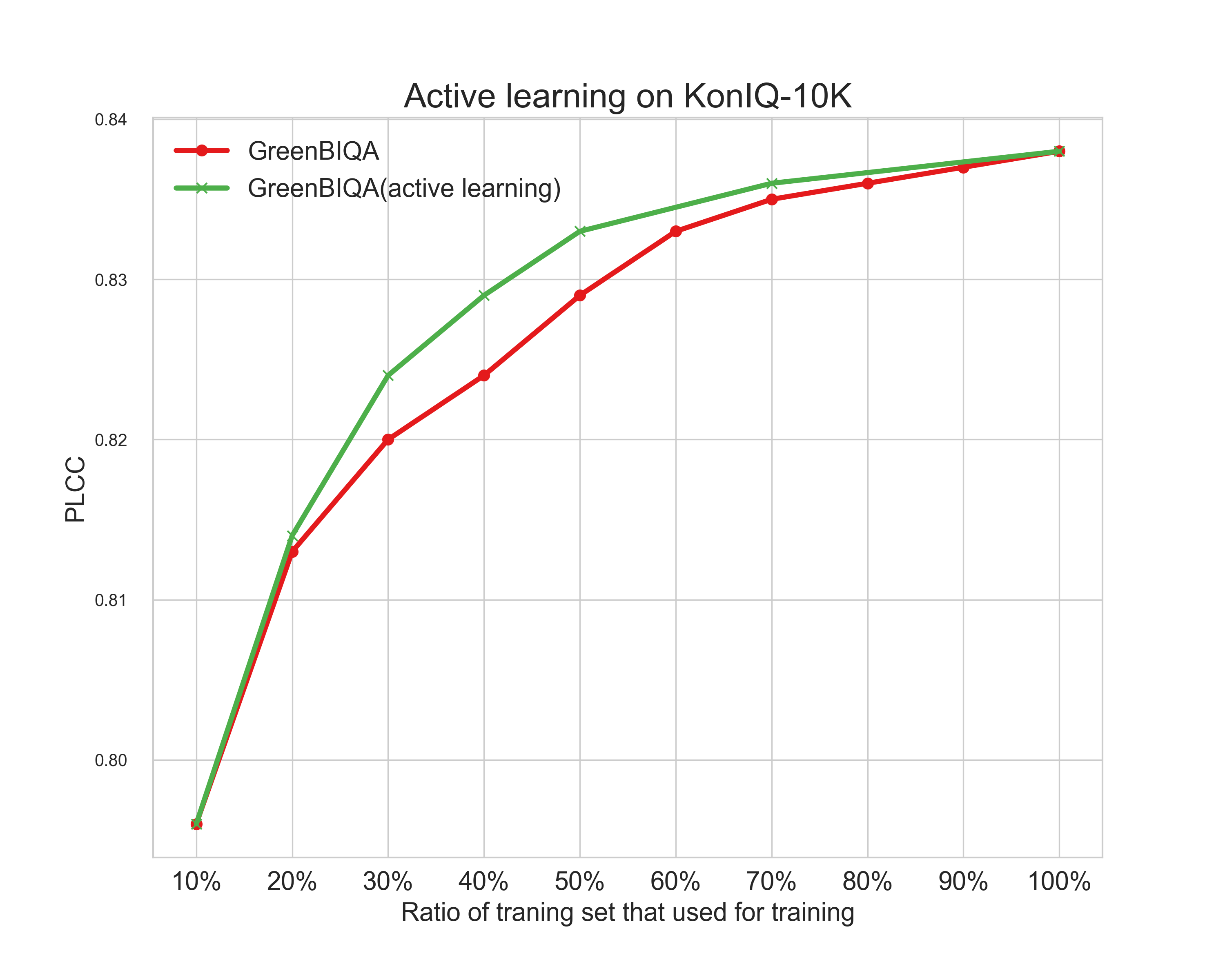}\\
\caption{Comparison of the PLCC performance of GreenBIQA using active
learning (in green) and random selection (in red) on the KonIQ-10k
dataset.}\label{fig:active_learning}
\end{figure}

\subsection{Weak Supervision}

We train BIQA models using different percentages of the KonIQ-10K
training dataset (e.g. from 1\% to 90\%) as shown in Fig.
\ref{fig:weak_learning} and show the PLCC performance against the full
test dataset.  For a fair comparison, we only compare GreenBIQA with
WaDIQaM, which is a simple DL method.  Note that we do not choose
advanced DL methods with pre-trained networks for performance
benchmarking since pre-trained networks have been trained by other
larger datasets. We show the mean and the plus/minus one standard
deviation. We see that GreenBIQA performs robustly under the weak
supervision setting. Even if it is only trained on 1\% of training samples,
GreenBIQA can achieve a PLCC value higher than 0.67.  Conversely,
WaDIQaM does not perform well when the percentage goes low since a small
number of samples is not sufficient in the training of a large neural
network.  

\subsection{Active Learning}

To further investigate the potential of GreenBIQA, we implement an
active learning scheme \cite{rouhsedaghat2021low,settles2009active} below. 
\begin{enumerate}
\item Keep the initial training set as 10\% of the full training dataset
and obtain an initial model denoted by $M_1$. 
\item Predict the performance of remaining samples in the training
dataset using $M_i$, $i=1,2, \cdots, 8$. Compute the standard derivation
of predicted scores of all sub-images associated with the same image,
which indicates prediction uncertainty.
\item Select a set of images that have the highest standard derivations
in Step 2, where its size is 10\% of the full training dataset.  Merge
them into the current training image set; namely, their ground truth
labels are leveraged to train Model $M_{i+1}$. 
\end{enumerate}
We repeat the above process in sequence to obtain models $M_1, \cdots,
M_9$.  Model $M_{10}$ is the same as the one that uses all training
samples.  We compare the PLCC performance of GreenBIQA with active
learning and with random sampling in Fig.  \ref{fig:active_learning}.
We see that the active learning strategy can improve the performance of
the random selection scheme in the range from 20\% to 70\% of full
training samples.

\begin{figure*}[!htbp]
     \centering
  \begin{tabular}{c@{}c@{}c@{}c}
    \includegraphics[width=.24\linewidth]{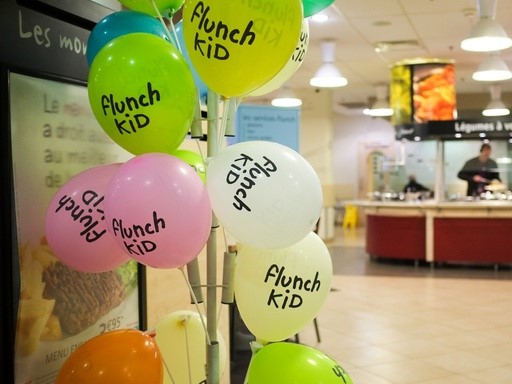} &
    \includegraphics[width=.24\linewidth]{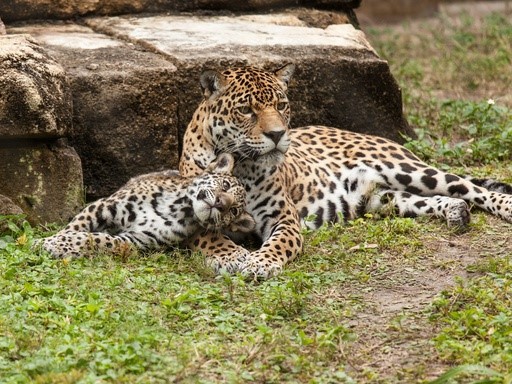} &
    \includegraphics[width=.24\linewidth]{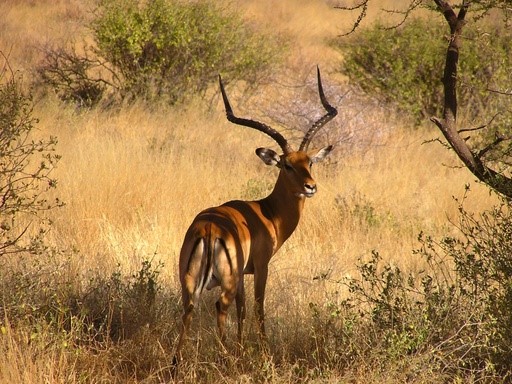} &
    \includegraphics[width=.24\linewidth]{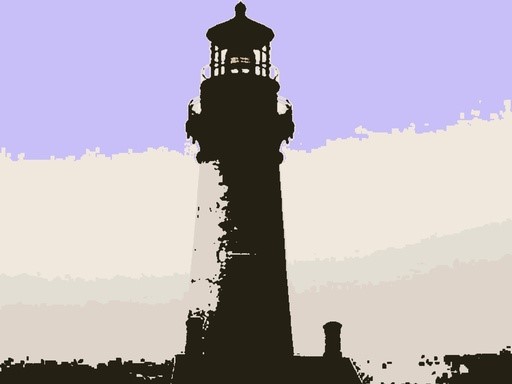} 
    \\[\abovecaptionskip]
    \small MOS(G) = 3.63 & \small MOS(G) = 3.72 & \small MOS(G) = 4.21 & \small MOS(G) = 1.78\\
    \small MOS(P) = 3.65 & \small MOS(P) = 3.81 & \small MOS(P) = 3.75 & \small MOS(P) = 2.68
  \end{tabular}
\caption{Comparison of the ground truth MOS and Green-BIQA predicted MOS values 
of four exemplary images, which are represented by MOS(G) and MOS(P), 
respectively.}\label{fig:examples}
\end{figure*}

\section{Conclusion and Future Work}\label{G_sec:conclusion}

A lightweight high-performance blind image quality assessment method,
called GreenBIQA, was presented in this paper.  Its PLCC and SROCC
performance was evaluated on two synthetic-distortion datasets and two
authentic-distortion datasets.  It outperforms all conventional BIQA
methods and simple DL methods in all four datasets.  As compared to the
state-of-the-art advanced DL methods with pre-trained models, GreenBIQA
still has the best performance in synthetic datasets and offers
close-to-best performance in authentic datasets. Above all, GreenBIQA
has tremendous advantages in its small model sizes, fast inference time,
and low computational complexities (in terms of FLOPs). These properties
make GreenBIQA an ideal BIQA choice in mobile and edge devices.

There are several research topics worth further investigation.  First,
we show four exemplary images with their ground truth and GreenBIQA
predicted MOS values in Fig. \ref{fig:examples}. The left two give good
prediction scores while the right two illustrate poor prediction scores.
GreenBIQA underestimates the MOS score of the long-horn deer image due
to the blurred background.  GreenBIQA overestimates the MOS score of
the light tower image due to lack of sufficient training samples of
similar characteristics. To further improve the performance of
GreenBIQA, we need a low-cost mechanism to find the attention region
and/or a way to check the prediction confidence.  Second, it is
important to generalize GreenBIQA to lightweight high-performance video
quality assessment by incorporating temporal information. 

\bibliographystyle{unsrt}  
\bibliography{references}

\end{document}